\documentclass[11pt,twoside]{article}

\usepackage{asp2006}
\usepackage{epsf}
\usepackage{psfig}
\usepackage{lscape}

\def\msun{\hbox{${\rm M}_{\odot}$}}

\begin{document}
\title{Magnetic topologies of cool stars}   
\author{J.-F.~Donati$^1$, M.M.~Jardine$^2$, P.~Petit$^1$, J.~Morin$^1$, J.~Bouvier$^3$, 
A.C.~Cameron$^2$, X.~Delfosse$^3$, B.~Dintrans$^1$, W.~Dobler$^4$, C.~Dougados$^3$, 
J.~Ferreira$^3$, T.~Forveille$^3$, S.G.~Gregory$^2$, T.~Harries$^5$, G.A.J.~Hussain$^2$, 
F.~M\'enard$^3$, F.~Paletou$^1$}   
\affil{$^1$~LATT, Obs.\ Midi-Pyr\'en\'ees, CNRS/UPS, 14 av.\ Belin, F-31400 Toulouse, France \\ 
$^2$~School of Physics and Astronomy, Univ.\ of St~Andrews, St~Andrews KY16 9SS, UK \\ 
$^3$~LAOG, Obs.\ de Grenoble, CNRS/UJF, BP 53, F-38041 Grenoble cedex 9, France \\ 
$^4$~Physics \& Astronomy, Univ.\ of Calgary, 2500 Univ.\ Dr, Calgary, Alberta T2N 1N4, Canada \\ 
$^5$~School of Physics, Univ.\ of Exeter, Stocker Road, Exeter EX4~4QL, UK}    

\begin{abstract} 
Stellar magnetic fields can be investigated using several, very
complementary approaches.  While conventional spectroscopy is capable
of estimating the average magnetic strength of potentially complex
field configurations thanks to its low sensitivity to the vector
properties of the field, spectropolarimetry can be used to map the
medium- and large-scale structure of magnetic topologies.  In
particular, the latter approach allows one to retrieve information
about the poloidal and toroidal components of the large-scale dynamo
fields in low-mass stars, and thus to investigate the physical
processes that produce them.  Similarly, this technique can be
used to explore how magnetic fields couple young stars to their
massive accretion disc and thus to estimate how much mass and 
angular momentum are transfered to the newly-born low-mass star.  
We present here the latest results in this field obtained with
spectropolarimetry, with special emphasis on the surprising discoveries 
obtained on very-low mass fully-convective stars and classical T Tauri stars 
thanks to the ESPaDOnS spectropolarimeter recently installed on 
the 3.6m Canada-France-Hawaii Telescope.
\end{abstract}


\section{Context}  

Most cool stars like our Sun are known to exhibit signs of activity.  These 
activity demonstrations can take different forms and occur on different 
timescales.  Cool spots are seen to come and go at the surfaces of cool stars 
\citep[for a recent review, see, eg,][]{Berdyugina05},  
and their number, in the particular case of the Sun, is seen to vary regularly 
on a period (dubbed activity cycle) of about 11~yrs.  Very-hot low-density 
plasma is observed around cool stars, forming the so-called corona that 
shows up very clearly around the Sun during total eclipses.  Flares often take 
place on cool stars and are usually associated with massive ejections of 
coronal material into the interplanetary/interstellar medium.  These activity 
demonstrations usually show up in stellar spectra through a large number of proxies, 
like for instance  X-ray luminosity or core emission in Ca~{\sc ii} H and K lines.  
The level of activity of cool stars usually scales up with their rotation rate.  

Our current understanding is that such activity demonstrations  
are a by-product of the complex and variable magnetic fields that cool stars 
generate at their surfaces and in their convective envelopes through 
phenomena called dynamo processes, involving cyclonic motions of ionised 
plasma and rotational shearing of internal layers.  In the particular case of 
the Sun, these processes are thought to 
take place in a thin interface layer confined at the base of the convective 
zone, where rotation gradients are supposed to be largest 
\citep[for recent reviews, see, eg,][]{Fan04, Charbonneau05}.  Very-low-mass 
stars, whose interiors are fully convective and thus obviously lack such an 
interface layer where dynamo processes could concentrate, are thus 
especially interesting for studies of stellar activity;  while such 
characteristics should make these stars unable to produce solar-like magnetic 
fields, they are nevertheless both very active and strongly magnetic 
\citep{Delfosse98, JohnsKrull96}, 
and the exact mechanism under which they yet succeed at producing their 
magnetism is still a very debated subject 
\citep[see, eg,][]{Dobler06, Chabrier06}.  

An even hotter issue concerns the role of magnetic fields in the formation of 
stars and their planetary systems.  During this evolutionary phase, the star 
is supposed to form from a collapsing molecular cloud, turned into a flat and 
massive accretion disc under the effect of both gravity and rotation.  Accretion 
in such discs is unusually strong, typically orders of magnitude stronger than what 
molecular viscosity can achieve.  These discs are often spatially associated 
with powerful and highly collimated jets emerging from the disc core and aligned 
with the disc rotation axis.  Through their spectral energy distributions and other 
tracers, we 
infer that these discs feature a central hole, suggesting that accretion from the 
inner disc ridge towards the central stars rather occurs through dense and 
discrete funnels.  In all three cases, models invoke magnetic fields as the main 
explanation \citep[for recent reviews, see, eg,][]{Pudritz06, Bouvier06};  
such fields are also expected to modify the rate at which 
protoplanetary clumps form and migrate within the disc.  While strong magnetic 
fields are indeed known to be present at the surfaces of such newly-born stars 
(dubbed classical T~Tauri stars or cTTS), very little is known on their exact 
topologies at the surfaces of both stars and discs.  This is however precisely 
the kind of information we need to constrain existing models.  

Magnetic fields on stars are most often estimated using the well-known Zeeman 
effect, that broadens the unpolarised profiles of magnetically sensitive lines 
and induces circular (Stokes V) and linear polarisation (Stokes Q and U) signals 
throughout their widths, 
depending in particular on the orientation of field lines with respect to the 
line of sight.  For this very reason, high-resolution stellar spectropolarimetry, 
that can recover information on large- and medium-scales magnetic topologies 
at the surfaces of active stars, is an obvious tool to use to challenge existing 
theoretical models of stellar formation and dynamo processes \citep[eg,][]{Donati03}.  
ESPaDOnS, the new 
generation spectropolarimeter recently installed on the 3.6m Canada-France-Hawaii 
telescope (CFHT), and NARVAL, its twin copy just commissionned on the 2m 
Bernard-Lyot telescope (TBL) atop Pic du Midi, were optimised in this aim and 
fill a long-standing gap in instrumental capabilities (Donati et al, in preparation).  
Recent surprising results 
obtained with ESPaDOnS in the above mentioned research fields are outlined in 
this paper.  Future potential directions for new instruments are also suggested.

\section{Measuring and modelling magnetic fields with spectropolarimetry}  

Until 1980, all experiments at measuring magnetic fields in cool active stars 
other than the Sun 
failed.  Most of them were using instruments directly inherited from solar physics, 
measuring line shifts between spectra respectively measured in circular left and 
right polarisation states and giving access to the average magnetic field component 
along the line of sight (dubbed longitudinal field).  However, for complex magnetic 
topologies such as that of the Sun and those anticipated on cool stars, the net 
longitudinal magnetic field is fairly small, with contributions of opposite 
polarities mutually cancelling out.  This a posteriori explains why these 
initial attempts failed.  

From 1980, investigations of the differential broadening of unpolarised spectral 
lines with different magnetic sensitivities demonstrated unambiguously that 
magnetic fields are indeed present at the surfaces of cool stars 
\citep[for a recent review, see][]{JohnsKrull07}.  This success was obtained thanks 
to the fact that broadening signatures from regions of opposite polarities do not 
mutually cancel in the integrated spectrum as polarisation signatures do.  
The very reason behind the success of this technique is however also what 
causes its intrinsic limitations.  Being almost insensitive to the local 
magnetic topology, this technique yields no more than an estimate of the relative 
surface area covered with magnetic fields, along with an average magnetic intensity 
(sometimes a rough distribution of magnetic intensities) within these magnetic 
regions.  

Since 1990, various studies demonstrated that polarisation signatures in spectral 
lines of cool stars, although often very weak, are actually detectable provided 
that full Zeeman signatures (rather than longitudinal fields values only) are 
recorded, and that specific and optimised instrument, observing procedure and 
reduction software are used \citep{Donati97}.  While this latter technique remains 
insensitive to the small magnetic scales present at the surface of cool stars, it 
can however yield key information such as how much fractional magnetic energy 
is stored within large and medium spatial scales, and how the field decomposes into 
its axisymmetric and non-axisymmetric modes, or into its poloidal and toroidal 
components.  In this respect, it provides us with a genuinely new and very powerful 
tool for studying dynamo processes of active stars other than the Sun.  

For such studies, we use high-resolution spectropolarimeters.  
They consist of an achromatic polarimeter mounted at the Cassegrain focus of a 
telescope, fiber feeding a bench-mounted high-resolution \'echelle spectrograph on 
which both orthogonal components of the selected polarisation state can be simultaneously 
recorded as interleaved \'echelle spectra on the CCD detector.  It also includes all 
usual calibration, viewing and guiding facilities that are necessary for conventional 
spectroscopy.  The new generation of such instruments, ESPaDOnS and NARVAL, were 
especially optimised to feature a very achromatic polarimetric analysis (using Fresnel 
rhombs as retarders), a very high efficiency (of order 10 to 15\% including telescope 
and detector), a wide spectral coverage (370 to 1,000~nm in a single exposure) and 
high spectral resolution (65,000 in polarimetric mode).  A complete instrumental 
description is given in Donati et al (in preparation).   

\begin{figure}[!t]
\centerline{\psfig{file=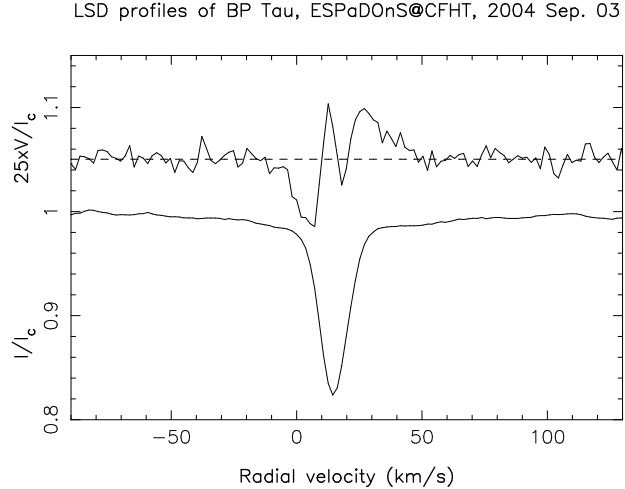,angle=270,width=8.4cm}}
\caption[]{LSD circular polarisation Zeeman signature from the photospheric lines 
of BP Tau, as derived from ESPaDOnS data}
\label{fig:lsd}
\end{figure}

Given that typical Zeeman signatures from cool active stars are rather small 
\citep{Donati97}, detecting them usually requires the extraction of the relevant information 
from as many lines as possible throughout the entire spectrum, using cross-correlation 
type tools such as Least-Squares Deconvolution \citep[LSD,][]{Donati97}.  Up to 8,000 
lines can be used in the domain of ESPaDOnS, yielding average Zeeman signatures with 
an equivalent signal to noise ratio boosted by several tens compared to that of a 
single average spectral line (see Fig.~\ref{fig:lsd}).    

Measuring Zeeman signatures of cool active stars and monitoring their modulation 
as the star rotates gives access, to some extent, to the parent magnetic topology at 
the surface of the stars, in the very same way medical imaging techniques recover 
3D images of the inside structure of human bodies by looking at them from all sides.  
Such tomographic techniques, dubbed Doppler and Zeeman-Doppler imaging, use the fact 
that surface features located at different longitudes and latitudes, and hosting field 
lines of different orientations produce different time-dependent distortions in 
temporal series of spectral lines \citep{DonatiBrown97}.  While such techniques are 
most efficient for stars rotating rapidly, they can also successfully be used on slow 
rotators \citep[eg,][]{Donati06b} and are particularly good at recovering 
both poloidal and toroidal components of the surface magnetic field.  
Assuming a potential field, the poloidal field component can also be used to estimate, 
through vector field extrapolation techniques, the large-scale coronal structure 
of the stellar magnetosphere \citep{Jardine02}.

\section{Very-low-mass fully-convective stars}  

As mentionned already, very-low-mass main-sequence stars (with a spectral type later 
than M3, ie a mass lower than about 0.35~\msun, \citealt{ChabrierBaraffe97}) are  
particularly interesting for 
studies of dynamo processes in active stars.  Being fully convective, these stars 
obviously lack the thin interface layer at the base of the convective zone where 
solar-like dynamo processes presumably concentrate.  Yet, they are both very active 
and strongly magnetic.  Current dynamo theories predict that they should trigger a 
genuine type of non-solar dynamo processes in which cyclonic convection and turbulence 
are by far the main actors, as opposed to the Sun where differential rotation is a 
major contributor.  The most recent investigations conclude that such stars should 
be able to produce large-scale magnetic fields;  however, while some claim that they 
should rotate as solid bodies and host purely non-axisymmetric large-scale field 
configurations \citep[eg,][]{Kuker05, Chabrier06}, others diagnose that they should still 
generate significant differential rotation and thus produce a net (though weak) 
axisymmetric field component \citep[eg,][]{Dobler06}.  

\begin{figure}[!t]
\centerline{\psfig{file=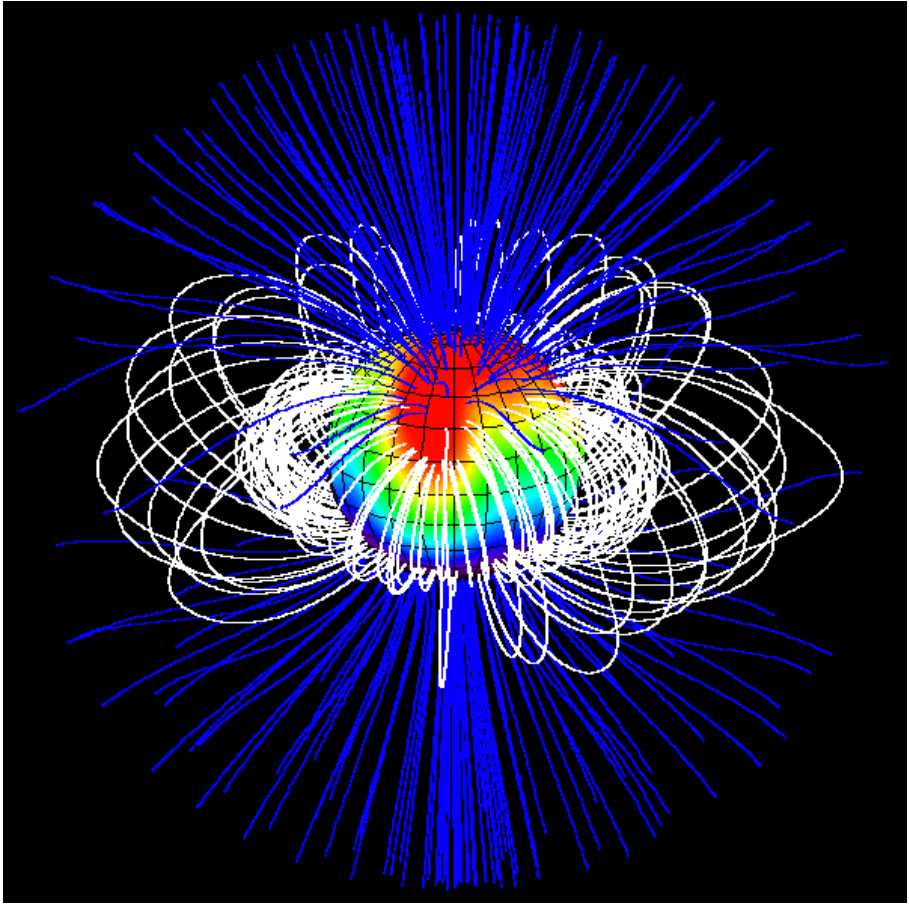,width=12cm}}
\caption[]{Magnetosphere of V374 Peg, as derived from ESPaDOnS spectropolarimetric data}
\label{fig:mdw}
\end{figure}

Observations of largely convective stars indicate that surface differential rotation 
is indeed vanishing with increasing convective depths \citep{Barnes05}, with fully 
convective stars rotating as solid bodies \citep{Donati06a}.  At first sight, this result 
seems to confirm nicely the theoretical predictions of \citet{Kuker05} and to invalidate 
those of \citet{Dobler06}.  However, spectropolarimetric observations of large-scale 
magnetic topologies in fully-convective stars led to a fairly different conclusion.  
By looking with ESPaDOnS at the rapidly rotating M4.5 dwarf V374~Peg (whose rotation 
period is just under half a day, ie about 60 times shorter than that of the Sun) for 
almost three complete rotation periods over nine rotation cycles, \citet{Donati06a} 
derived that the star hosts a strong poloidal field whose energy mostly concentrates 
in low-degree axisymmetric modes (see Fig.~\ref{fig:mdw}), 
in gross contradiction with all existing predictions.  Also worth noting is the drastic 
difference between the field configuration of V374~Peg with those of partly 
convective stars, which always feature a significant (and often dominant) toroidal
component \citep[][]{Donati03, Petit05}.  The main conclusion of these first observations 
is that fully-convective stars indeed seem to be able to trigger different kinds of 
dynamo processes than those of partly convective stars.  There is however still some way 
to go before we fully understand the theoretical machinery of dynamo processes in stellar 
convective zones, and it may well be that the solar dynamo itself is more complex than 
what we think it is.  

Following this initial result, a more extensive spectropolarimetric observing program of 
fully-convective dwarfs was initiated, in the aim of achieving a small magnetic survey of 
about 25 stars selected to sample as evenly as possible rotation rates and spectral types.  
Up to now, observations have been collected for about half a dozen stars, with spectral 
types spanning M3 to M7 and periods ranging from half a day up to about 5~d.  Although 
still very incomplete, this survey already suggests that among fully convective stars, 
strong axisymmetric large-scale field topologies are found exclusively on rapid rotators 
while non-axisymmetric magnetic structures only show up on moderate rotators, the slowest 
rotators showing no large-scale fields at all (Morin et al., in preparation).  This 
result, once confirmed on a more comprehensive data set, could stress how critically 
rotation influences dynamo processes and hint at new ideas for producing a more successful 
theoretical description of how dynamo works in stellar convective zones.

\section{Classical T-Tauri stars and their accretion discs}  

As briefly alluded to above, magnetic fields play a key role in the early stages of 
stellar evolution when the newly born star is still surrounded by a massive accretion 
disc (the cTTS phase).  In most cases, the spectral energy distribution of such objects 
is dominated by light from the star in the visible, while the accretion disc outshines 
the star in the infrared;  in rare cases though, the accretion disc is found to undergo 
drastic outbursts (called FU~Ori episodes) during which the rate of accretion scales 
up by several orders of magnitude, causing the disc to outshine the star at all 
wavelengths.  In protostellar discs, magnetic fields are expected to boost the accretion 
rate within the disc itself (through MHD instabilities), to expell 
some of the disc material along the jet-like 
structures often observed around young stars, and to clear out by magnetic disruption 
the central regions of the disc and funnel the inner disc material to the stellar 
surface (when the accretion rate is not too strong).  

Obviously, magnetic fields are expected to have a strong impact on the angular momentum 
evolution of young stars.  Theorists proposed that the magnetic coupling between the 
stars and their inner discs was strong enough to force the star to corotate at the 
Keplerian velocity of the inner disc \citep[eg,][]{Konigl91}, thus potentially explaining 
why rotation rates of cTTS are significantly lower than those of discless protostars.  
Yet, estimates of magnetic intensities at the surface of cTTS derived from Zeeman 
broadening of unpolarised lines is largely uncorrelated with those models predict to 
ensure star-disc coupling \citep{JohnsKrull07}.  It essentially shows that the physical 
details of this magnetic link between the star and disc are still rather enigmatic to 
us, and need to be unravelled if we want to understand angular momentum evolution during 
stellar formation stages.  

In this respect, spectropolarimetric observations of cTTS can bring us new clues for 
this problem.  The discovery that spectral lines formed at the base of accretion funnels 
(eg the He~{\sc i} line at 587.6~nm) exhibit strong levels of circular polarisation 
\citep{JohnsKrull99} indicate that magnetic fields are indeed likely to control  
accretion from the inner disc to the stellar surface.  
Further observations collected across the rotation cycles of a few cTTS \citep{Valenti04, 
Symington05} revealed that the large-scale magnetic geometry, anchored 
in kG field regions, seems rather simple and stable on time scales of at least several 
years, suggesting the presence of strong dipoles moderately tilted with respect 
to the stellar rotation axis.  
Numerical simulations of magnetospheric accretion were carried out in this context, 
including various additional ingredients such as winds from the star and disc and 
dynamo disc fields \citep[eg,][]{Romanova03, Romanova04, vonRekowski04}, to obtain 
more realistic predictions along with detailed observational diagnostics.  

The apparent weakness of Zeeman signatures in photospheric lines of cTTs 
\citep[eg,][]{JohnsKrull99} remains however mysterious in the context of strong 
dipole-like magnetospheric topologies.  
Prominent circular polarisation signals are indeed expected in most lines 
from strong dipole fields, while only highly tangled magnetic topologies can remain 
undetected through spectropolarimetry.  

With the availability of ESPaDOnS, new perspectives are opened for studying magnetospheric 
accretion on cTTS.  The K7 low-mass cTTS BP~Tau was monitored for 9 consecutive 
nights in Feb.~2006 while the more massive G5 cTTS V2129~Oph was looked at in June~2005 for 
8 successive nights.  In both of them, strong Zeeman signatures from numerous emission 
lines formed in accretion funnel footpoints were detected and found to vary smoothly and 
simply with rotation rate, in complete agreement with previous results \citep{Valenti04, 
Symington05}.  
In addition to this, weaker (though still very clear) Zeeman signatures were also detected 
in photospheric lines thanks to the higher efficiency of ESPaDOnS (eg, see Fig.~\ref{fig:lsd}); 
their compex shape however confirms that the surface field topology on both BP~Tau and 
V2129~Oph is indeed more complex than a simple dipole.  

\begin{figure}[!t]
\centerline{\psfig{file=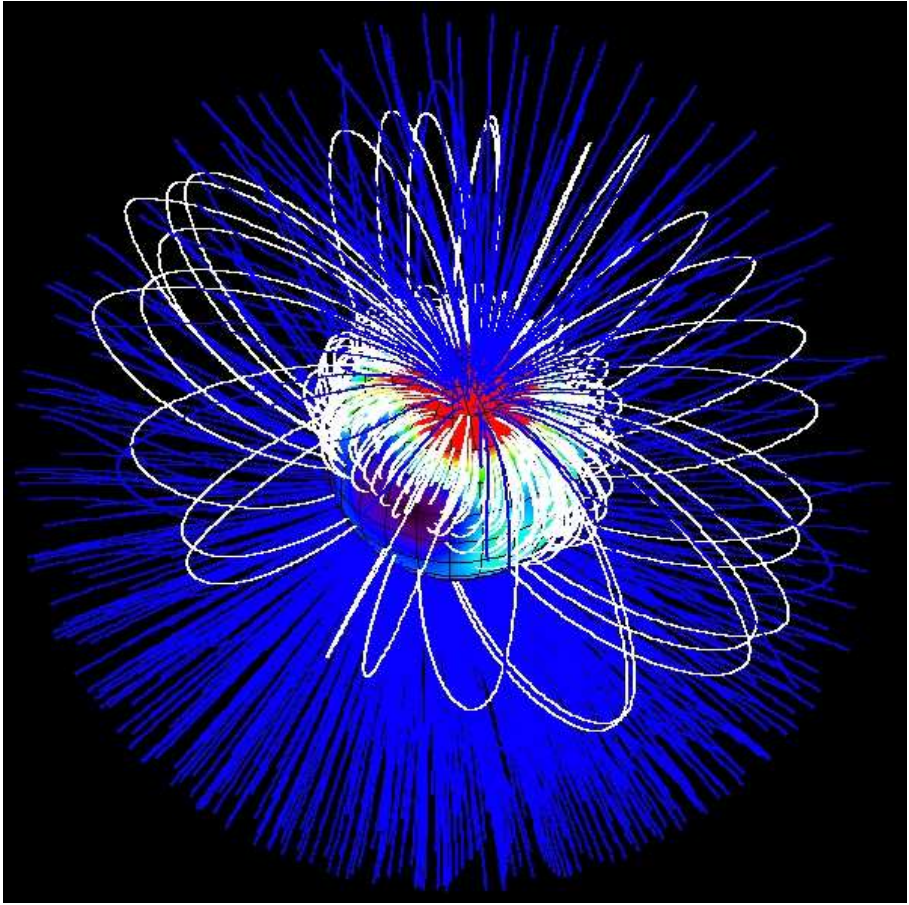,width=12cm}}
\caption[]{Magnetosphere of V2129 Oph, as derived from ESPaDOnS spectropolarimetric data 
(this result is still preliminary)}
\label{fig:ctt}
\end{figure}

Attempts at finding a unique magnetic model that can reconcile both sets of Zeeman 
signatures are being carried out at the moment, under the assumption that photospheric lines 
are not sensitive to the highly magnetic (and thus presumably very cool) surface regions 
located at the footpoints of accretion funnels.    
A preliminary solution for V2129~Oph (see Fig.~\ref{fig:ctt}) illustrates what the true 
magnetosphere of a cTTS could look like (Donati et al., in preparation);  
while the large-scale field indeed 
resembles a dipole as witnessed from the simple variational behaviour of the 
Zeeman signatures from accretion proxies, the surface field is more complex and 
features nearby field regions of opposite polarities.  
More realistic simulations 
of magnetospheric accretion are thus needed (and underway, \citealt{Jardine06, Gregory06a, 
Gregory06b}) to obtain a better understanding of this crucial 
stage in stellar formation.  

Protostellar accretion discs may also host magnetic fields.  In fact, they are the 
likely vectors that convey interstellar magnetic fields within stellar cores through 
ambipolar diffusion and advection.  Magnetic fields are probably playing a main role 
in boosting accretion rates through instabilities such as the now famous 
magneto-rotational instability \citep[eg,][]{Balbus03} - they are also the major 
ingredient with which accretion discs succeed at producing powerful collimated 
jets and expell a significant fraction of their mass and angular momentum  
\citep[eg,][]{Pudritz06}.  

Magnetic fields are known to be present within molecular cores \citep[eg,][]{Girart06} 
and in the external regions of accretion discs \citep[eg,][]{Huta03}, suggesting that 
interstellar magnetic fields are indeed playing a role in stellar 
formation.  However, until very recently, no observation existed 
about the putative magnetic field in the inner regions of accretion discs where 
the jets are actually fired and the surface densities culminate.   The new results 
obtained by ESPaDOnS on FU~Ori demonstrate that strong disc core magnetic fields 
are indeed present, with topologies compatible with what is predicted by collimated jet 
formation theories \citep{Donati05}.  These observations, although still very 
fragmentary, also indicate that the magnetic plasma within the disc is apparently 
slowed down more that what models predict. 

Extensive spectropolarimetric monitoring of protostellar accretion discs such as 
FU~Ori carried over several Keplerian periods (for disc radii up to a few 0.1~AU), 
should allow mapping, through Doppler tomography, the full distribution of 
densities and magnetic fields with radius and azimuth throughout the disc core.  
In particular, it could tell us how successful theoretical models of 
magnetised molecular cloud core collapse \citep[eg,][]{Banerjee06} are.  It 
could also give us a completely new opportunity to investigate the formation 
and migration of protoplanetary clumps within cores of accretion discs as well 
as the associated role of magnetic fields, and thus test the 
various theoretical models yet proposed in the literature \citep[eg,][]{Terquem03}.

\section{Conclusions and prospects}  

Our knowledge of magnetic topologies of cool active stars has greatly improved 
in the last two decades thanks to spectropolarimetry.  The recent introduction 
of new generation instruments with increased sensitivity (such as 
ESPaDOnS at CFHT) allowed to explore new classes of objects such as young 
forming stars and fully convective dwarfs.  As a result, we are now able to detect 
magnetic topologies in most cool stars provided that their rotation and activity 
are high enough and that they host a significant large-scale field component.  

When coupled to tomographic techniques, spectropolarimetry allows not only to 
detect, but also to characterise the geometry of stellar magnetic topologies.  
For instance, we are able to decompose the field topology into its poloidal 
and toroidal components, thus offering great promises for studies of dynamo 
processes in cool stars other than the Sun.  Such studies have already revealed 
major surprises, the latest being that fully-convective stars are apparently able 
to trigger strong axisymmetric large-scale field components, against most 
theoretical expectations.  Similarly, extended observations of young stars 
should soon allow in-depth testing of most theoretical models attempting to 
describe stellar formation in the presence of magnetic fields.  

These results suggest that high-resolution spectropolarimetry should be actively 
developed in the coming years, with efficient ESPaDOnS-like instruments routinely 
installed on intermediate size instruments offering reasonable access for the 
long-term monitoring runs that such studies require.  At the same time, new 
instruments should be designed to expand further our spectropolarimetric 
capabilities and access new types of diagnostics;  in this respect a high 
resolution near-infrared spectropolarimeter providing full spectral coverage 
from 0.9 to 2.4~microns appears as a very promising option for the study of 
magnetic fields in cool stars.

\acknowledgements 

JFD thanks the SOC of CS14 and CNRS for providing financial support for 
attending the conference.



\begin{thebibliography}{}
\bibitem[Balbus \& Hawley(2003)]{Balbus03}
Balbus S.A.\ \& Hawley J.F., 2003, LNP 614, 329 
\bibitem[Banerjee \& Pudritz(2006)]{Banerjee06}
Banerjee R.\ \& Pudritz R.E., 2006, \apj\ 641, 949 
\bibitem[Barnes et al.(2005)]{Barnes05}
Barnes J.R., Cameron A.C., Donati J.-F., James D.J., Marsden S.C., 
Petit P., 2005, \mnras\ 357, L1  
\bibitem[Berdyugina(2005)]{Berdyugina05}
Berdyugina S., 2005, Living Reviews in Solar Physics \# 8 
\bibitem[Bouvier et al.(2006)]{Bouvier06}
Bouvier J., Alencar S.H.P., Harries T.J., Johns-Krull C.M., Romanova M.M., 2006, in 
Protostars and Planets V, ed.\ B.\ Reipurth, D.\ Jewitt, \& K.\ Keil 
\bibitem[Chabrier \& Baraffe(1997)]{ChabrierBaraffe97}
Chabrier G.\ \& Baraffe I., 1997, \aap\ 327, 1039 
\bibitem[Chabrier \& K\"uker(2006)]{Chabrier06}
Chabrier G.\ \& K\"uker M., 2006, \aap\ 446, 1027 
\bibitem[Charbonneau(2005)]{Charbonneau05}
Charbonneau P., 2005, Living Reviews in Solar Physics \# 2 
\bibitem[Delfosse et al.(1998)]{Delfosse98}
Delfosse X., Forveille T., Perrier C., Mayor M., 1998, \aap\ 331, 581 
\bibitem[Dobler, Stix, \& Brandenburg(2006)]{Dobler06}
Dobler W., Stix M., \& Brandenburg A., 2006, \apj\ 638, 336 
\bibitem[Donati \& Brown(1997)]{DonatiBrown97}
Donati J.-F.\ \& Brown S.F., 1997, \aap\ 326, 1135 
\bibitem[Donati et al.(1997)]{Donati97}
Donati J.-F., Semel M., Carter B.D., Rees D.E., Cameron A.C., 2003, \mnras\ 291, 658 
\bibitem[Donati et al.(2003)]{Donati03}
Donati J.-F.\ et al., 2003, \mnras\ 345, 1145 
\bibitem[Donati et al.(2005)]{Donati05}
Donati J.-F., Paletou F., Bouvier J., Ferreira J., 2005, Nature 438, 466  
\bibitem[Donati et al.(2006a)]{Donati06a}
Donati J.-F., Forveille T., Cameron A.C., Barnes J.R., Delfosse X., 
Jardine M.M., Valenti J.A., 2006a, Science 311, 633 
\bibitem[Donati et al.(2006b)]{Donati06b}
Donati J.-F.\ et al., 2006b, \mnras\ 370, 629 
\bibitem[Fan(2004)]{Fan04}
Fan Y., 2004, Living Reviews in Solar Physics \# 1 
\bibitem[Girart, Rao, \& Marrone(2006)]{Girart06}
Girart J.M., Rao R., \& Marrone D.P., 2006, Science 313, 812 
\bibitem[Gregory et al.(2006a)]{Gregory06a}
Gregory S.G., Jardine M.M., Simpson I., Donati J.-F., 2006a, \mnras\ 371, 999
\bibitem[Gregory et al.(2006b)]{Gregory06b}
Gregory S.G., Jardine M.M., Cameron A.C., Donati J.-F., 2006b, \mnras\ 373, 827
\bibitem[Hutawarakorn \& Cohen(2003)]{Huta03}
Hutawarakorn B.\ \& Cohen R.J., 2003, \mnras\ 345, 175  
\bibitem[Jardine et al.(2006)]{Jardine06}
Jardine M.M., Cameron A.C., Donati J.-F., Gregory S.G., Wood K., 2006, \mnras\ 367, 917 
\bibitem[Jardine et al.(2002)]{Jardine02}
Jardine M.M., Wood K., Cameron A.C., Donati J.-F., Mackay D.H., 2002, \mnras\ 336, 1364 
\bibitem[Johns-Krull \& Valenti(1996)]{JohnsKrull96}
Johns-Krull C.M.\ \& Valenti J.A., 1996, \apj\ 459, L95 
\bibitem[Johns-Krull et al.(1999)]{JohnsKrull99}
Johns-Krull C.M., Valenti J.A., Hatzes A.P., Kanaan A., 1999, \apj\ 510, L41  
\bibitem[Johns-Krull(2007)]{JohnsKrull07}
Johns-Krull C.M., 2007, these proceedings
\bibitem[K\"onigl(1991)]{Konigl91}
K\"onigl A., 1991, \apj\ 370, L39  
\bibitem[K\"uker \& R\"udiger(2005)]{Kuker05}
K\"uker M.\ \& R\"udiger G., 2005, AN 326, 265 
\bibitem[Petit et al.(2005)]{Petit05}
Petit P., et al., 2005, \mnras\ 361, 837 
\bibitem[Pudritz et al.(2006)]{Pudritz06}
Pudritz R.E., Ouyed R., Fendt C., Brandenburg A., 2006, in 
Protostars and Planets V, ed.\ B.\ Reipurth, D.\ Jewitt, \& K.\ Keil 
\bibitem[Romanova et al.(2003)]{Romanova03}
Romanova M.M., Ustyugova G.V., Koldoba A.V., Wick J.V., Lovelace R.V.E., 2003, \apj\ 595, 1009 
\bibitem[Romanova et al.(2004)]{Romanova04}
Romanova M.M., Ustyugova G.V., Koldoba A.V., Lovelace R.V.E., 2004, \apj\ 610, 920 
\bibitem[Symington et al.(2005)]{Symington05}
Symington N., Harries T., Kurosawa R., Naylor T., 2005, \mnras\ 358, 977 
\bibitem[Terquem(2003)]{Terquem03}
Terquem C.E.J.M.L.J., 2003, \mnras\ 341, 1157 
\bibitem[Valenti \& Johns-Krull(2004)]{Valenti04}
Valenti J.A.\ \& Johns-Krull C.M., 2004, \apss\ 292, 619  
\bibitem[vonRekowski \& Brandenburg(2004)]{vonRekowski04}
von Rekowski B.\ \& Brandenburg A., 2004, \aap\ 420, 17   
\end{thebibliography}
\end{document}